\documentclass[preprintnumbers, amsmath, amssymb, showpacs,twocolumn, superscriptaddress, 
nofootinbib]{revtex4}

\usepackage{graphicx,subfigure,bm,color,psfrag,hyperref}
\usepackage{amsfonts}
\usepackage{lipsum}
\usepackage{mathtools}
\usepackage{verbatim}

\newcommand{\be}{\begin{equation}}
\newcommand{\ee}{\end{equation}}
\newcommand{\bea}{\begin{eqnarray}} 
\newcommand{\eea}{\end{eqnarray}}
\newcommand{\beaa}{\begin{eqnarray*}}
\newcommand{\eeaa}{\end{eqnarray*}}

\renewcommand{\)}{\right)}

\renewcommand{\]}{\right]}

\begin{document}

\title{Modified cosmology from extended entropy with varying exponent}

 \author{Shin'ichi Nojiri}
\email{nojiri@gravity.phys.nagoya-u.ac.jp}
\affiliation{Department of Physics, Nagoya University, Nagoya 464-8602, Japan}
\affiliation{Kobayashi-Maskawa Institute for the Origin of Particles and the
Universe, Nagoya University, Nagoya 464-8602, Japan}

\author{Sergei D. Odintsov}
\email{odintsov@ieec.uab.es}
\affiliation{Institut de Ciencies de lEspai (IEEC-CSIC), 
Campus UAB, Carrer de Can Magrans, s/n 
08193 Cerdanyola del Valles, Barcelona, Spain}
\affiliation{Instituci\'{o} Catalana de Recerca i Estudis Avan\c{c}ats
(ICREA), Passeig Llu\'{i}s Companys, 23 08010 Barcelona, Spain}

\author{Emmanuel N. Saridakis}
\email{Emmanuel\_Saridakis@baylor.edu}
\affiliation{Department of Physics, National Technical University of Athens, Zografou
Campus GR 157 73, Athens, Greece}
\affiliation{Department of Astronomy, School of Physical Sciences, University of Science 
and Technology of China, Hefei 230026, P.R. China}

\begin{abstract}
We present a modified cosmological scenario that arises from the application of 
non-extensive thermodynamics with varying exponent. We extract the modified Friedmann 
equations, which contain new terms quantified by the non-extensive exponent, 
possessing standard $\Lambda$CDM cosmology as a subcase. 
Concerning the universe evolution at late times we obtain an effective dark energy 
sector, and we show that we can acquire the usual thermal history, with the successive 
sequence of matter and dark-energy epochs, with the effective 
dark-energy equation-of-state parameter being in the quintessence or in the phantom 
regime. The interesting feature of the scenario is that the above 
behaviors can be obtained even if the explicit cosmological constant is set to zero, 
namely they arise purely from the extra terms.
Additionally, we confront the model with Supernovae type Ia and Hubble parameter 
observational data, and we show that the agreement is very good.
Concerning the early-time universe we obtain inflationary de~Sitter 
solutions, which are driven by an effective cosmological constant that includes 
the new terms of non-extensive 
thermodynamics. This effective screening can provide a
description of both inflation and late-time acceleration with the same parameter choices,
which is a significant advantage. 

\end{abstract}


\maketitle
\section{Introduction}
 
According to the Standard Model of Cosmology, the universe experienced two phase of 
accelerating expansion, one at early (inflation) and one at late times. In principle 
there are two ways to explain these behaviors. The first is to maintain general 
relativity as the gravitational theory and introduce new energy contents, such 
as the dark energy sector \cite{Peebles:2002gy,Cai:2009zp} and the 
inflaton field \cite{Bartolo:2004if}. The second is to assume that the 
extra degrees of freedom that drive the universe acceleration arise from a modified 
theory of gravity (for reviews see 
\cite{Nojiri:2003ft,Nojiri:2006ri,Capozziello:2011et,Cai:2015emx}). 
 Concerning modified gravities, the simplest models are $F(R)$ gravity 
\cite{Capozziello:2002rd,Nojiri:2006gh,Nojiri:2010wj}, $f(G)$ 
gravity \cite{Nojiri:2005jg}, Weyl gravity
\cite{Mannheim:1988dj, Flanagan:2006ra}, Galileon theory 
\cite{Nicolis:2008in, Deffayet:2009wt, Leon:2012mt} etc. However, one could follow more 
radical ways of modification, i.e. start from the torsion instead of the curvature 
gravitational formulation and construct 
$f(T)$ gravity \cite{Ben09,Chen:2010va}, $f(T,T_G)$ gravity
\cite{Kofinas:2014owa,Kofinas:2014daa}, etc.

However, a different approach to modified gravity arises from the connection between 
gravity and thermodynamics \cite{Jacobson:1995ab,Padmanabhan:2003gd,Padmanabhan:2009vy}.
In particular, it is known that in a cosmological framework one can express the Friedmann 
equations as the first law of thermodynamics applied in the 
universe apparent horizon \cite{Cai:2005ra,Akbar:2006kj,Cai:2006rs}, and 
equivalently he can apply the 
first law of thermodynamics in the universe horizon and result to 
the Friedmann equations. The above conjecture about the ``thermodynamics of space-time'' 
\cite{Jacobson:1995ab} has been applied in various classes of modified gravity 
\cite{Cai:2006rs,Akbar:2006er,Paranjape:2006ca,Sheykhi:2007zp,Jamil:2009eb,Cai:2009ph,
Wang:2009zv,Jamil:2010di, Gim:2014nba, Fan:2014ala}. 

Recently, there have appeared some works in the literature in which the above 
thermodynamical consideration is applied using extended entropy relations
instead of the usual one 
\cite{Tsallis:2012js,Komatsu:2013qia,Nunes:2014jra,Lymperis:2018iuz,Saridakis:2018unr, 
Sheykhi:2018dpn,Artymowski:2018pyg,Abreu:2017hiy,Jawad:2018frc,Zadeh:2018wub, 
daSilva:2018ehn}. In particular, it is known that in the case of non-additive systems, 
such as gravitational ones, the standard Boltzmann-Gibbs additive entropy should be 
generalized to the non-extensive Tsallis entropy 
\cite{Tsallis:1987eu,Lyra:1998wz,Wilk:1999dr}, 
which can be applied in all cases, 
possessing the former as a limit. The non-extensivity is 
parametrized by a new exponent $\delta$, with the value 1 corresponding to the standard 
entropy. Hence, the cosmological application of this 
non-extensive thermodynamics results to new modified Friedmann equations that possess the 
usual ones 
as a particular limit, namely when the Tsallis generalized entropy becomes the usual one, 
but which in the general case contain extra terms that appear for the first time.

In the present work we are interested in investigating the extended case in which the 
exponent of the non-extensive thermodynamics has a running behavior,
namely that it varies according to the energy scale.
Such a running behavior is known to be the typical case for quantum field theory and 
quantum gravity when renormalization group is applied. In this case, the coupling 
constants (even cosmological and gravitational coupling constants) are running with the 
energy scale. When such theories are applied in a cosmological framework it turns 
out that the running is with   time. Hence, our proposal to consider a running 
behavior of non-extensive thermodynamics arises from quantum field 
theoretical considerations, which although not envisioned by Tsallis when he constructed  
his approach, are in principle necessary when ones tries to embed Tsallis entropy into a 
general framework that would be consistent with quantum gravity  setup.

In particular,   entropy corresponds to the physical 
degrees of freedom of a system, however the renormalization of a quantum theory implies 
that the degrees of freedom depend on the scale. 
In   standard field theory, in low energy regime, massive modes decouple and therefore 
the degrees of freedom decrease. In   gravity case the situation becomes more
complicated, and if the space-time fluctuations become large in the 
ultraviolet regime then the degrees of freedom may increase. On the other hand, if 
gravity 
becomes topological, the degrees of freedom will decrease, which 
could be consistent with holography. 

From the above discussion we may conclude that in both high and low 
scales the exponent $\delta$ may acquire values away from the standard value 1, while at 
intermediate scales it should be close to unity. Therefore, the cosmological application 
of such a running non-extensive thermodynamics will bring qualitatively new extra terms in 
the modified Friedmann equations, that are expected to play a role both at high-energy 
scales (inflation) as well as at low ones (late-time universe). In the following we will 
study in detail such a cosmological scenario. 

The plan of the work is the following. In 
Section \ref{model0} we review the relation of cosmology with thermodynamics. In Section 
\ref{model1} we apply non-extensive thermodynamics with varying exponent in a 
cosmological framework and we extract the modified Friedmann equations. Additionally, we 
investigate the scenario at late and early times, describing both dark energy and 
inflationary solutions. In Section \ref{fRcorr} we examine the possible correspondence of 
the scenario at hand with $F(R)$ gravity. Finally, in Section \ref{Conclusions} we 
summarize our results.

 \section{Cosmology from thermodynamics}
 \label{model0}
 
In this section we present the cosmological application of thermodynamical 
considerations.
Throughout the manuscript we work with a homogeneous and isotropic flat
Friedmann-Robertson-Walker (FRW) geometry with metric
\begin{equation}
ds^2=-dt^2+a^2(t)\,\delta_{ij} dx^i dx^j\, ,
\label{metric}
\end{equation}
where $a(t)$ is the scale factor. In the following subsections we analyze the case of 
standard and non-extensive thermodynamics separately.

\subsection{Standard thermodynamics}
\label{standtherm}
 
 We start by considering the expanding universe filled with 
a perfect fluid, with energy density $\rho$ and pressure $p$. Although it is not 
straightforward to determine the ``volume'' of the above system, namely to find 
the ``radius'' that forms its boundary, in the literature there is a consensus that 
this should be the apparent horizon 
\cite{Cai:2005ra,Cai:2008gw}, which in the case of a flat universe becomes
\begin{equation}
\label{apphor}
r_H=\frac{1}{H},
\end{equation}
with $H=\frac{\dot a}{a}$ the Hubble parameter and with dots 
denoting derivatives with respect to $t$ (hence in a flat three-dimensional geometry the 
apparent horizon coincides with the Hubble one).
The apparent horizon is a marginally trapped surface with vanishing expansion
\cite{Bak:1999hd}, and in the case of dynamical space-times it corresponds to a
causal horizon associated with the gravitational entropy and the surface gravity
\cite{Bak:1999hd,Hayward:1997jp,Hayward:1998ee}. 

Let us now investigate the thermodynamics of the system bounded by $r_H$. 
The energy going outwards through the horizon,  in time interval $dt$, is given by 
\cite{Cai:2005ra}
\begin{equation}
\label{Tslls2}
dQ = - dE = \frac{4\pi}{3} r_H^3 \dot\rho dt = \frac{4\pi}{3H^3} \dot\rho dt \, ,
\end{equation}
with $ \frac{4\pi}{3} r_H^3$ the system's volume. 
By using the standard conservation law, namely
\begin{equation}
\label{Tslls3}
0 = \dot \rho + 3 H \left( \rho + p \right) \, ,
\end{equation}
then Eq.~(\ref{Tslls2}) can be rewritten as 
\begin{equation}
\label{Tslls4}
dQ = \frac{4\pi}{H^2} \left( \rho + p \right) dt \, .
\end{equation}
At this stage we should attribute to the universe horizon a temperature and an entropy.
Taking into account the black hole temperature and entropy relations, one deduces that 
the corresponding temperature is the Hawking temperature 
\cite{Cai:2005ra,Cai:2008gw}
\begin{equation}
\label{Tslls6}
T = \frac{1}{2\pi r_H} = \frac{H}{2\pi}\,,
\end{equation}
while the entropy relation is the usual Bekenstein-Hawking 
relation \cite{Padmanabhan:2009vy}
\begin{equation}
\label{Tslls5}
S = \frac{A}{4G}\, ,
\end{equation}
in units where $\hbar=k_B = c = 1$, with 
$A = 4\pi r_H^2 = \frac{4\pi}{H^2} $ the horizon area and $G$ the 
gravitational constant.
Hence, inserting (\ref{Tslls4}),(\ref{Tslls6}) and (\ref{Tslls5}), into the first law of 
thermodynamics
\begin{equation}
\label{Tslls6B}
TdS = dQ \, ,
\end{equation}
we obtain 
\begin{equation}
\label{Tslls7}
\dot H = - 4\pi G \left( \rho + p \right) \, ,
\end{equation}
which is nothing else than the second Friedmann equation. Finally, integrating
Eq.~(\ref{Tslls7}) and using the 
conservation law (\ref{Tslls2}), we obtain 
the 
first FRW equation, namely
\begin{equation}
\label{Tslls8}
H^2 = \frac{8\pi G}{3} \rho + \frac{\Lambda}{3} \,,
\end{equation}
where $\Lambda$ is an integration constant that plays the role of the cosmological 
constant. 

\subsection{Non-extensive thermodynamics}

Let us now apply the above procedure, but instead of the standard entropy 
relation we use the generalized, non-extensive, Tsallis entropy. As we mentioned in the 
Introduction, in systems with diverging partition function, such as large-scale 
gravitational systems, the standard 
Boltzmann-Gibbs theory cannot be applied. In these cases one needs to use 
non-extensive, 
Tsallis thermodynamics, which still possesses standard 
Boltzmann-Gibbs theory as a limit. Thus, the standard 
Boltzmann-Gibbs additive entropy needs to be generalized to the non-extensive, 
non-additive entropy, namely Tsallis entropy 
\cite{Tsallis:1987eu,Lyra:1998wz,Wilk:1999dr,Nunes:2014jra}, which in units where 
$\hbar=k_B = c = 1$ can be written in compact form as 
\cite{Tsallis:2012js}:
\begin{equation}
\label{Tslls9}
S = \frac{A_0}{4 G} \left(\frac{A}{A_0} \right)^\delta\, .
\end{equation}
In the above expression $A$ is the the area of the system, $A_0$ is a 
constant introduced for dimensional reasons, and $\delta$ 
is the new parameter that quantifies the non-extensivity. In the case where 
$\delta=1$ one re-obtains the standard 
Bekenstein-Hawking entropy.
 
We repeat the steps of the previous subsection, namely we apply 
the first law of thermodynamics (\ref{Tslls6B}) in the universe apparent horizon 
(\ref{apphor}), with (\ref{Tslls4}) and 
(\ref{Tslls6}), however concerning the entropy we use the non-extensive relation 
(\ref{Tslls9}). In this case 
we obtain
\begin{equation}
\label{Tslls10}
\delta \left( \frac{H_1^2}{H^2} \right)^{\delta -1} \dot H 
= - 4\pi G \left( \rho + p \right) \, ,
\end{equation}
where for convenience we have introduced the constant $H_1$ through 
$A_0 \equiv \frac{4\pi}{H_1^2}$.
Finally, integrating (\ref{Tslls10}) we result to 
\cite{Lymperis:2018iuz}
\begin{equation}
\label{Tslls11}
\frac{\delta}{2 - \delta} H_1^2\left( \frac{H^2}{H_1^2} \right)^{2 - \delta}
= \frac{8\pi G}{3} \rho + \frac{\Lambda}{3} \, ,
\end{equation}
with $\Lambda$ an integration constant. Equation (\ref{Tslls11}) is the generalized 
Friedmann equation arising from non-extensive horizon thermodynamics, and its novel extra 
terms can be used to describe either an effective dark energy sector or the inflation 
realization \cite{Lymperis:2018iuz}.

Before proceeding to the next Section, where we will extend the above framework, let us 
make an interesting comment on the equivalence between the modified cosmology through 
non-extensive thermodynamics and the model of holographic dark energy \cite{Li:2004rb}.
The latter consideration is based on the holographic principle 
\cite{Fischler:1998st}, which when it is applied in a cosmological setup it leads to a 
dark 
energy density of the form 
\begin{equation}
\label{HH1}
\rho_\Lambda = \frac{3c^2}{8\pi G L^2}\, ,
\end{equation}
where $L$ is the infrared cutoff of the theory and $c$ the model parameter. Concerning 
the 
choice of $L$, this could be the particle horizon $L_p = a \int_0^t \frac{dt}{a}$, the 
future event horizon $L_f = a \int_t^\infty \frac{dt}{a}$, or the Ricci scalar $R$,
however one could use a more general cutoff $L_g$ which is a function of 
$L_p$, $L_f$, $R$, as well as of the cosmological constant 
$\Lambda=\frac{12}{l^2}$ \cite{Nojiri:2005pu}, namely 
\begin{equation}
\label{Tslls19}
L_g = L_g \left( L_p, L_f, R, l \right) \, .
\end{equation}
If we consider the choice
\begin{equation}
\label{Tsll19B}
\frac{1}{L_g^2} = \frac{1}{c^2R^2} - \frac{3}{8\pi G l^2} 
 - \frac{\beta}{l^{2\alpha} L_p^{2(1-\alpha)}} \, ,
\end{equation}
with $\alpha$ and $\beta$ constants, insert that in the holographic energy density 
(\ref{HH1}) and then in the usual Friedmann equation
\begin{equation}
\label{Tslls19C}
H^2 = \frac{8\pi G}{3}\left( \rho + \rho_\Lambda \right) + \frac{\Lambda}{3} \, ,
\end{equation}
we obtain
\begin{equation}
\label{Tslls19D}
\frac{8\pi G\beta}{3 l^{2\alpha} L_p^{2(1-\alpha)}} 
= \frac{8\pi G}{3} \rho \, .
\end{equation}
In the case where the scale factor has a power-law evolution $a\propto 
t^\gamma$, we find $L_p \propto t$ and 
$H\propto t^{-1}$, and therefore the left hand side of (\ref{Tslls19D}) behaves 
as $H^{2\left(1-\alpha\right)}$. Thus, if identify $\alpha - 1=2 - \delta$, this specific 
model of generalized holographic dark energy reproduces the modified cosmology 
(\ref{Tslls11}) arising from non-extensive horizon thermodynamics.

\section{Modified cosmology from non-extensive entropy with varying exponent}
\label{model1}

In this section we investigate the modified cosmology that arises from thermodynamical 
considerations, applying the non-extensive entropy relation but allowing it to have 
a running behavior, namely that it can vary with the energy scale. As we mentioned in the 
Introduction, the entropy corresponds to physical 
degrees of freedom, but the renormalization of a quantum theory implies 
that the degrees of freedom depend on the scale. In the gravitational case, if the 
space-time fluctuations become large in the 
ultraviolet regime then the degrees of freedom may increase, while if 
gravity becomes topological the degrees of freedom may decrease. Hence, we conclude 
that in general the exponent $\delta$ of Tsallis entropy (\ref{Tslls9}) can have a 
running 
behavior.

In the cosmological framework the energy scale can be quantified by the value of the 
Hubble parameter $H$. Hence, in the following we assume that $\delta$ 
has the scale-dependence $\delta \equiv\delta(x) $, with
$x=\frac{H_1^2}{H^2} $, and where $H_1$ is a parameter with units of $H$ that sets the 
reference scale. Repeating 
the procedure of the previous section, that is applying
the first law of thermodynamics (\ref{Tslls6B}) in the universe apparent horizon 
(\ref{apphor}), with (\ref{Tslls4}),
(\ref{Tslls6}), and (\ref{Tslls9}), and with a varying $\delta$ we result to 
\begin{equation}
\label{Tslls16}
\left\{ \delta 
+ \left[ \frac{H_1^2}{H^2} \ln \left( \frac{H_1^2}{H^2} \right) \right]
\delta' 
\right\} \left( \frac{H_1^2}{H^2} \right)^{\delta -1} \dot H 
= - 4\pi G \left( \rho + p \right) \, ,
\end{equation}
where $\delta'(x)\equiv\partial \delta(x)/\partial x$ (from now on a prime denotes the 
derivative of a function with respect to its argument). Thus, integrating (\ref{Tslls16}) 
and using (\ref{Tslls3}) we find
\begin{equation}
\label{Tslls17}
\left. - H_1^2 \left\{ x^{\delta(x) - 2} + 2 \int^x dx x^{\delta(x) -3} \right\} 
\right|_{x=\frac{H_1^2}{H^2}}= \frac{8\pi G}{3} \rho + \frac{\Lambda}{3} \, .
\end{equation}
Equation (\ref{Tslls17}) is the modified Friedmann equation that arises from 
non-extensive 
thermodynamics with varying exponent, and one of the main results of the present work. In 
the following we will study its cosmological implications in detail.

In order to proceed we need to consider a specific ansatz for $\delta(x)$. In principle 
one can make various choices and elaborate Eq.~(\ref{Tslls17}) numerically. However, 
since we desire to obtain analytical solutions we will choose forms that allow it. 
Additionally, from the class of choices that allow for analytical solutions we will focus 
on the $\delta(x)$-forms that present the physical behavior mentioned in the 
Introduction. In particular, as we described, in both high 
and low scales we expect $\delta(x)$ to acquire values away from the standard value 1, 
while at intermediate scales it should be close to unity. Hence, a general class of 
$\delta(x)$ that exhibits this behavior and simultaneously allows for analytical 
solutions of the integral in (\ref{Tslls17}) is 
 \begin{equation}
\label{Tslls17Bdelta}
\delta(x) = \frac{ \ln \left[c \left( x^{3-n} + \alpha(x) b_2 x^{2-n} + b_1 b_2^2 
x^{1-n}\right)
\right] }{\ln x}\,,
\end{equation}
where 
\begin{equation}
\label{Tslls17Balpha}
\alpha (x)\equiv \frac{n (3 - n)}{(1-n)^2} + \frac{n^2}{(1+n)(2-n)} b_1 b_2^2 
x^{-n-1}\,, 
\end{equation}
with $n$, $b_1$, $b_2$ the model parameters and 
$ c \equiv \left\{ \frac{3 - n}{(1-n)^2} + \frac{b_1}{1+n} 
\right\}^{-1}b_2^{n-2} $. In this case, Eq.~(\ref{Tslls17}) becomes 
\begin{equation}
\label{Tslls17Ca}
\left. - f(x) \right|_{x=\frac{H_1^2}{H^2}} H_1^2
= \frac{8\pi G}{3} \rho + \frac{\Lambda}{3}\, ,
\end{equation}
with 
\begin{align}
\label{Tslls17Cb}
f(x) \equiv c & \left[ \left( \frac{3 - n}{1-n}\right) x^{1-n} 
 -\left( \frac{2 - n}{n}\right) b_2 \alpha(x) x^{-n}\right.\nonumber\\
&\ \left. -\left( \frac{1-n}{1+n}\right) b_1 b_2^2 x^{-n-1} \right] \, .
\end{align}
Note that when $n=2$ and $b_1=b_2=0$ we obtain $\delta(x)=1$, i.e., standard 
thermodynamics, 
and in this case 
Eq.~(\ref{Tslls17Ca}) gives the standard Friedmann equation, namely
Eq.~(\ref{Tslls8}).

\subsection{Late-time universe}

Let us now investigate the modified Friedmann equations from non-extensive 
thermodynamics with varying exponent at late cosmological 
times. In this case, we can define an effective dark energy sector that includes all 
 the extra terms that non-extensive 
thermodynamics with varying exponent brings. In particular, we can re-write 
Eqs.~(\ref{Tslls17Ca}),(\ref{Tslls16})  as
\begin{align}
\label{Fr1b}
H^2 =& \frac{8\pi G}{3}(\rho + \rho_\mathrm{DE}) \\
\dot{H} =& -4\pi G(\rho +p+\rho_\mathrm{DE}+p_\mathrm{DE})\, ,
\label{Fr2b}
\end{align}
where 
\begin{align}
\label{rhode}
\rho_\mathrm{DE}=& \frac{3}{8\pi G}\left[
\frac{\Lambda}{3}+\left. H_1^2 f(x) \right|_{x=\frac{H_1^2}{H^2}}+H^2
\right] \, ,\\
\label{pde}
p_\mathrm{DE}=&-\frac{1}{8\pi G}\Big\{
\Lambda+H_1^2
\Bigl[
f(x)-\frac{2H_1^2\dot{H}}{H^4}f'(x)
\Bigr] \Bigr|_{x=\frac{H_1^2}{H^2}}
\nonumber\\
& \qquad \ \ \ \ \ \, +2\dot{H}+3H^2
\Big\}
\end{align}
are the energy density and pressure of the effective dark-energy density respectively. 
As one can verify using (\ref{rhode}), (\ref{pde}) the effective dark energy is 
conserved, since it satisfies
\begin{equation}\label{rhodeconserv}
\dot{\rho}_\mathrm{DE}+3H \left(\rho_\mathrm{DE}+p_\mathrm{DE} \right)=0\, .
\end{equation}
Furthermore, we can define the dark-energy equation-of-state parameter as
\begin{equation}\label{rhodeconserv}
w_\mathrm{DE}\equiv \frac{p_\mathrm{DE}}{\rho_\mathrm{DE}}\, .
\end{equation}
Additionally, 
it proves convenient to introduce the dark energy and matter density parameters through
\begin{align}
\Omega_\mathrm{m}\equiv&\frac{8\pi G}{3H^2}\rho \, , 
\label{Omm}\\
\Omega_\mathrm{DE}\equiv&\frac{8\pi G}{3H^2}\rho_\mathrm{DE}\, .
\label{ODE}
\end{align}
Finally, we introduce the deceleration 
parameter $q$ through
\begin{equation}
\label{qdeccel}
q\equiv-1-\frac{\dot{H}}{H^2}=\frac{1}{2}+\frac{3}{2}\left(w\Omega_m
+w_\mathrm{DE}\Omega_\mathrm{DE}  \right)\, ,
\end{equation}
where $w=p/\rho$ is the matter equation-of-state parameter.
 In summary, in the constructed modified cosmological scenario we can describe the 
late-time universe with the equations 
(\ref{Fr1b}) and (\ref{Fr2b}), as long as the matter equation-of-state parameter 
 is known. 

We proceed by numerically elaborating Eqs.~(\ref{Fr1b}),~(\ref{Fr2b}), focusing 
on the evolution of the observable quantities such as the density parameters and the 
dark-energy equation-of-state parameter. Additionally, in order to examine the 
capabilities of the model at hand in driving universe acceleration, we set the explicit 
cosmological constant $\Lambda$ to 0. 
For convenience, in the following as the 
independent variable we use the redshift $z$, defined through $1+z=a_0/a$,
and we set the current value of the scale factor to $a_0=1$. 
Moreover, we impose 
$\Omega_\mathrm{DE}(z=0)\equiv\Omega_\mathrm{DE0}\approx0.7$ as required 
by observations \cite{Ade:2015xua}.

\begin{figure}[ht]
\includegraphics[scale=0.46]{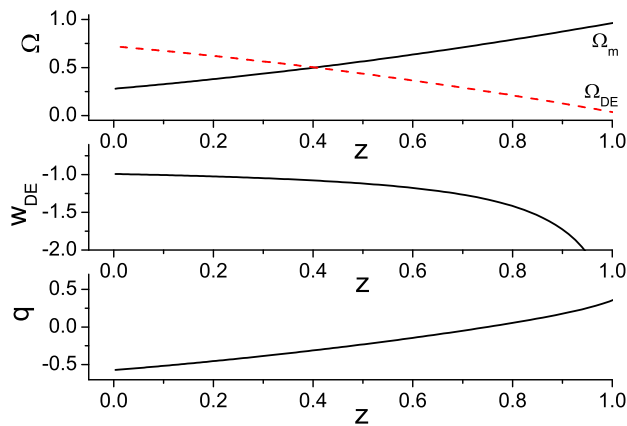}
\caption{
{\it{ Upper graph: The evolution of the effective dark energy 
density parameter $\Omega_\mathrm{DE}$ (black-solid) and of the matter density
parameter $\Omega_\mathrm{m}$ (red-dashed), as a function of the redshift $z$,
for the scenario of modified cosmology through non-extensive thermodynamics with varying 
exponent, for the parameter choices $\Lambda=0$, $b_1=0$, $n=4$, $H_1=1.1$ and $b_2=1$, 
in units where $8\pi G=1$. 
Middle graph: The evolution of the corresponding dark-energy equation-of-state parameter 
$w_\mathrm{DE}$. Lower graph: The evolution of the 
corresponding deceleration parameter $q$. In all graphs we have imposed 
 $\Omega_\mathrm{DE}(z=0)\equiv\Omega_\mathrm{DE0}\approx0.7$ 
at present in agreement with observations.
}} }
\label{Fig1}
\end{figure}

In the upper graph of Fig.~\ref{Fig1} we depict 
$\Omega_\mathrm{DE}$ and $\Omega_\mathrm{m}$ as a function of redshift, 
for the case of dust matter 
($w=0$), and with the parameter choices $\Lambda=0$, $b_1=0$, $n=4$, $H_1=1.1$ and 
$b_2=1$, in units where $8\pi G=1$. In the middle graph we draw the 
corresponding behavior of $w_\mathrm{DE}$, and in the 
lower graph we present the deceleration parameter.

From the upper graph of Fig.~\ref{Fig1} we can see that we obtain the usual 
thermal history of the universe, namely the successive sequence of matter and dark-energy 
epochs. Moreover, from the third graph of Fig.~\ref{Fig1} we observe that the 
transition from deceleration to acceleration happens at $z\approx 0.6$, in agreement with
observations. Finally, from the middle graph of Fig.~\ref{Fig1} we see that the 
value of $w_\mathrm{DE}$ at present is around $-1$, as required by observations, while in 
the 
past it may lie either in the quintessence or in the phantom regime.
We stress here that the above behaviors are obtained without the use of an explicit 
cosmological constant, that is they arise purely from the extra terms that modified 
cosmology through non-extensive thermodynamics with varying 
exponent brings. This is an advantage of the scenario showing the enhanced capabilities.

We close this subsection by providing simplifying analytical expressions for the energy 
density and pressure of the effective dark energy sector, namely (\ref{rhode}) and 
(\ref{pde}). Since we are considering the late-time universe we may focus on the regime 
$H^2\ll H_1^2$, i.e. $x\gg 1$. Hence, in this case (\ref{rhode}), (\ref{pde}) 
approximately give
\begin{align}
\label{rhode00}
\rho_\mathrm{DE} =& \frac{3}{8\pi G}\left[
\frac{\Lambda}{3} 
 - c\left( \frac{3 - n}{n-1} \right) \left( \frac{H_1^2}{H^2} \right)^{2-n} H^2
+H^2 \right] \, , \\
\label{pde00}
p_\mathrm{DE} =& -\frac{1}{8\pi G}\Big\{
\Lambda+2\dot{H}+3H^2\nonumber\\
& - c\Big( \frac{3 - n}{n-1} \Big)\Big( \frac{H_1^2}{H^2} \Big)^{2-n}
\Big[
H^2+2\dot{H}(n-1)
\Big]
\Big\},
\end{align}
and thus the first Friedmann equation (\ref{Fr1b}) becomes
\begin{equation}
\label{Tslls17F}
 c\left( \frac{3 - n}{n-1} \right) \left( \frac{H_1^2}{H^2} \right)^{2-n} H^2
= \frac{8\pi G}{3} \rho + \frac{\Lambda}{3} \, .
\end{equation}

We close the analysis of the late-time universe by  confronting the scenario with 
data from Supernovae type Ia (SNIa)  observations and direct $H(z)$ Hubble data,
extracting the  constraints on the model   parameters  using the maximum 
likelihood analysis. This can be obtained 
by minimizing the $\chi^2$ function in terms of the free parameters of the model $a_m$, 
assuming Gaussian errors, and applying the  Markov Chain
Monte Carlo (MCMC) algorithm within the Python package emcee
\cite{ForemanMackey:2012ig}.  The statistical vector 
of the free parameters is $a_m = (\Omega_{m0},\Lambda,n,b_2,H_1)$ (we restrict to 
$b_1=0$ for simplicity and we set the current value of the Hubble parameter $H_0$ to 
its Planck best-fit value $H_0\approx6\times10^{-61}$ (in units of $8\pi G=\hbar=c=1$) 
\cite{Ade:2015xua}). Hence,  the total $\chi^2$ of our datasets will be $  \chi^2_{tot} = 
\chi^2_{SN}+\chi^2_{H} $, where the separate $\chi^2$ are calculated as follows.

Concerning SNIa data   one measures the apparent luminosity in terms  of 
redshift, or equivalently the apparent 
magnitude. Therefore, we have
  \begin{equation}
  \chi ^2 _{SN} = \mu C_{SN,cov}^{-1} \mu^{T},
  \end{equation}
    where $\mu = \{ \mu_{\rm obs} - \mu_{\rm th}(z_{1}; a_m),..,\mu_{\rm obs} - \mu_{\rm 
th}(z_{N}; 
a_m) \} $ and $N=40$.  In the above expression
$\mu_{\rm obs}$ denotes the  observed distance modulus, which is
defined as the difference between the Supernova's absolute and 
apparent magnitude. We use the binned SNIa
data, as well as the corresponding  inverse covariance matrix  $C_{SN,cov}^{-1}$ from 
\cite{Scolnic:2017caz}. On the other hand, the theoretically calculated 
distance modulus $\mu_{\rm th}$  depends on the model parameters
$a_m$  through
\begin{equation}\mu_{\rm th}\left(z\right)=42.38-5\log_{10}h+5\log_{10}\[D_{L}
\left(z;a_m\right)\],
  \end{equation}
where   $D_{L}(z;a_m)$ is the dimensionless luminosity distance, reading as
  \begin{equation} D_{L}\(z;a_m\)\equiv\left(1+z\right)
\int^{z}_{0}dz'\frac{H_1}{H\left(z';a_m\right)}.
  \end{equation}
  The quantity $H\left(z';a_m\right)$ in the 
scenario at hand is obtained numerically from 
(\ref{Fr1b}),(\ref{rhode}), since it cannot be calculated analytically.
  \begin{figure}[ht]
  \includegraphics[width=0.34\textwidth]{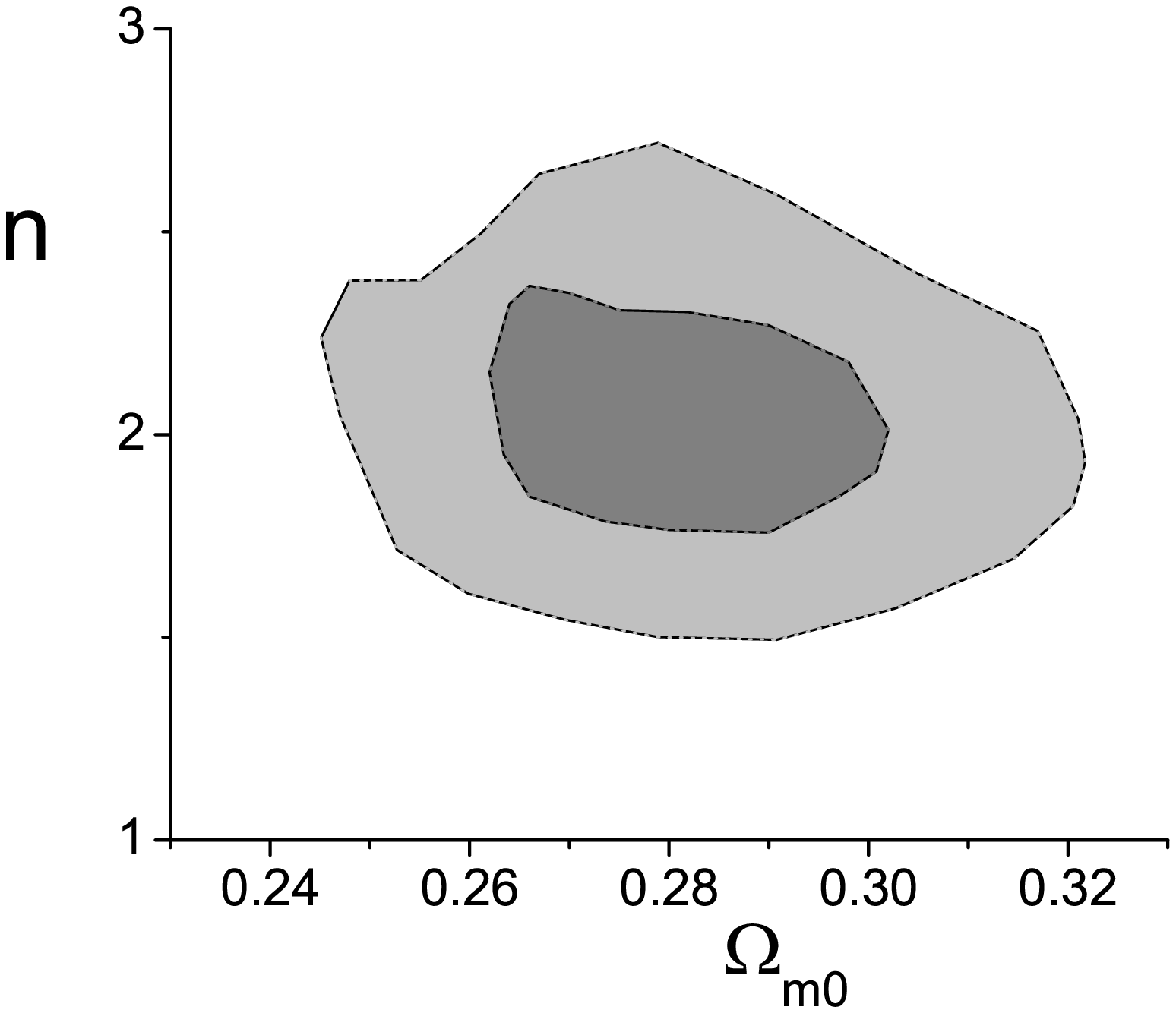}\\
  \vspace{-1.cm}
   \includegraphics[width=0.36\textwidth]{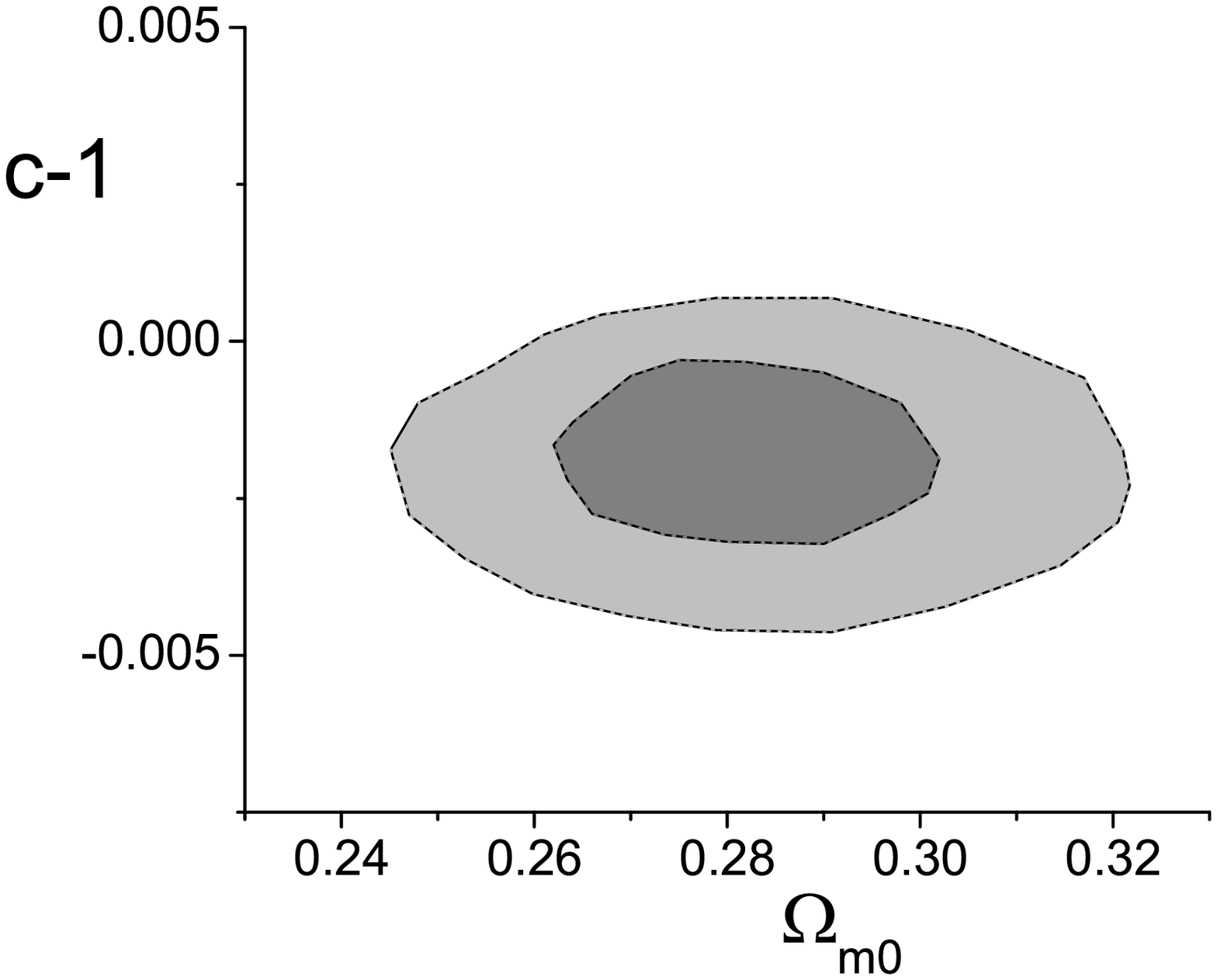}
\caption{{\it{The $1\sigma$ and $2\sigma$  2-dimensional contour 
plots for the parameters $n$ (upper graph) and $c$ (lower graph) in terms of the current 
value of the matter density parameter $\Omega_{m0}$, for the scenario of modified   
cosmology through non-extensive thermodynamics with varying exponent, using SNIa and 
$H(z)$ data. }} }
\label{cnuplot}
\end{figure}

Concerning the  direct measurements of the Hubble constant  
 we use the recent   data  from \cite{Yu:2017iju}, which include    $N=36$ measurements 
of 
$H(z)$ in  the range $0.07\leq z\leq 2.33$.  The corresponding $\chi^2$ is calculated as
\begin{equation}
\chi^{2}_{H}\left(a_m\right)={\cal H}\,
C_{H,\text{cov}}^{-1}\,{\cal H}^{T}\,,
\end{equation}
with ${\cal H}=\{H_{(1)}-H_{0}E(z_{1},a_m)\,,\,...\,,\,
H_{(N)}-H_{0}E(z_{N},a_m)\}$, $H_{(i)}$   the observed Hubble values at 
redshifts $z_{i}$ ($i=1,...,N$), and where  $C$ is the  
involved covariance matrix   
\cite{Nunes:2016qyp,Nunes:2016drj,Anagnostopoulos:2017iao,Basilakos:2018arq}. 
Finally, 
 the theoretical quantity $E(z_{i},a_m)\equiv 
H(z_{i},a_m)/H_0$  is obtained numerically from 
(\ref{Fr1b}),(\ref{rhode}).

In Fig. \ref{cnuplot} we provide the  contour 
plots for the parameters $n$ and $c$ in terms of the current 
value of the matter density parameter $\Omega_{m0}$, for the scenario of 
modified   cosmology through non-extensive thermodynamics with varying exponent, using 
SNIa and $H(z)$ data.  As we can see the agreement with the data is very good, and the 
matter energy density  coincides with that of Planck within 
1$\sigma$ \cite{Ade:2015xua}. For the new   
parameters of the scenario at hand we observe that they acquire values near their 
$\Lambda$CDM ones, namely $n=2$ and $b_1=b_2=0$ (i.e. $c=1$). Nevertheless, $c$ is away  
from its $\Lambda$CDM value at 1$\sigma$ confidence level, which shows that a slight 
deviation might be favored, however $\Lambda$CDM paradigm is included within  2$\sigma$.
Additionally, we mention that if we restrict our fittings to the case of $\Lambda=0$, 
then we acquire $n=3.56^{+1.02}_{-1.47}$ at 1 $\sigma$,    however with a large 
$\chi^2$. Finally, we would like to mention that the incorporation of  Cosmic Microwave 
Background (CMB) data, 
although necessary, requires a highly non-trivial treatment of the   $H(z)$ form, which 
in the model at hand in general cannot be obtained analytically. Such a 
detailed  elaboration lies beyond the scope of the present analysis. 

\subsection{Early-time universe}

In this subsection we study the modified Friedmann equation from 
non-extensive 
thermodynamics with varying exponent, namely Eq.~(\ref{Tslls17Ca}), at early times, 
namely we desire to examine the inflationary realization.
In this regime we may use the approximation $H^2\gg H_1^2$, i.e. $x\ll 1$. Thus, in the 
case $b_1\neq 0$
Eq.~(\ref{Tslls17Ca}) becomes
\begin{equation}
\label{Tslls17G}
c b_1 \left(\frac{1-n}{1+n}\right) \left( \frac{H_1^2}{H^2} \right)^{-n} H^2
= \frac{8\pi G}{3} \rho + \frac{\Lambda}{3} \, ,
\end{equation}
while in the case $b_1=0$ and $b_2\neq0$, we find
\begin{equation}
\label{Tslls17H}
 (2-n) b_2^{n-1} \left( \frac{H_1^2}{H^2} \right)^{1-n} H^2
= \frac{8\pi G}{3} \rho + \frac{\Lambda}{3}\, .
\end{equation}
Hence, neglecting the matter sector, Eqs.~(\ref{Tslls17G}),(\ref{Tslls17H}) give rise to
inflationary de~Sitter solutions. In particular, for $b_1\neq 0$ we obtain
\begin{equation}
\label{Tslls17M}
H = \mathrm{const.} = \left[ \frac{(1+n) \Lambda H_1^{2n}}{3(1-n) cb_1 } 
\right]^{\frac{1}{2(n+1)}}\, ,
\end{equation}
while for $b_1=0$ and $b_2\neq0$ we find
\begin{equation}
\label{Tslls17HH}
H = \mathrm{const.}= \left[ \frac{ \Lambda H_1^{2(n-1)}}{3 (2-n) b_2^{n-1}} 
\right]^{\frac{1}{2n}} \,.
\end{equation}

As we see, the scenario at hand accept inflationary de~Sitter solutions at early times. 
In particular, from (\ref{Tslls17M}),(\ref{Tslls17HH}) we may define an effective 
cosmological constant, namely 
\begin{equation}
\label{Tslls17M2}
\Lambda_\mathrm{eff}\equiv 3 \left[ \frac{(1+n) \Lambda H_1^{2n}}{3(1-n) cb_1 } 
\right]^{\frac{1}{(n+1)}} 
\end{equation}
for $b_1\neq 0$, and 
\begin{equation}
\label{Tslls17HH2}
\Lambda_\mathrm{eff}\equiv 3\left[ \frac{ \Lambda H_1^{2(n-1)}}{3 (2-n) b_2^{n-1}} 
\right]^{\frac{1}{n}} 
\end{equation}
for $b_1=0$, $b_2\neq0$.
Hence, the interesting feature of these solutions is that the cosmological 
constant is effectively screened, and the effective cosmological constant that drives 
inflation includes additionally the information of the new terms of non-extensive 
thermodynamics. This feature can be a great advantage in providing a 
description of both inflation and late-time acceleration with the same parameter choices, 
since the above effective screening gives the necessary enhanced acceleration in 
inflation comparing to dark-energy epoch.

As an illustrative example we consider the case $b_1=0$. 
Then, in the early universe relation (\ref{Tslls17HH}) gives $H^2 \sim 
\Lambda^{\frac{1}{n}} H_1^{\frac{2(n-1)}{n}}b_2^{\frac{1-n}{n}}$, while at late times 
relation (\ref{Tslls17F}) results to 
$H^2\sim \Lambda^{ \frac{1}{n-1}} H_1^{\frac{n-2}{n-1}}b_2^{\frac{2-n}{n-1}}$. 
Hence, knowing that 
in the late-time universe $H^2 \sim \left( 10^{-33}\, \mathrm{eV} \right)^2$ 
while in the early universe
$H^2 \sim \left( 10^{24}\, \mathrm{eV} \right)^2$,
we find that e.g. for $n=3/2$ and $H_1\sim b_2\sim1 \mathrm{eV} $ and with 
$\Lambda \sim 10^{-66}\,\mathrm{eV}^2$, both regimes can be obtained simultaneously.

\section{$F(R)$ gravity correspondence}
\label{fRcorr}

In this section we investigate the correspondence of modified cosmology through 
non-extensive thermodynamics with othe classes of modified gravity, and in particular 
with 
$F(R)$ gravity. As we showed in subsection \ref{standtherm}, starting from the first 
law of thermodynamics and using the standard 
Bekenstein-Hawking entropy, one can result to the standard Friedmann equations. On the 
other hand, the use of non-extensive Tsallis entropy leads to modified Friedmann 
equations. 
Hence, the question is whether there is a correspondence of these modified Friedmann 
equations with the modified Friedmann equations arising from modified gravity theories 
such as $F(R)$ gravity.

The action of $F(R)$ gravity, alongside the matter sector, reads as
\begin{equation}
\label{F1}
I = \frac{1}{16\pi G} \int d^4x \sqrt{-g} F(R) + I_m\, ,
\end{equation}
where $F(R)$ is a function of the scalar curvature ${R}$ and $I_m$ 
is the matter action. Extracting the field equations and applying them in FRW geometry 
we obtain the corresponding Friedmann equations, namely \cite{Nojiri:2006gh}
\begin{align}
\label{JGRG15}
0 =& -F(R) + 6\left(H^2 + \dot H\right) F'({R}) 
\nonumber\\
& - 36 \left( 4H^2 \dot H + H \ddot H\right) F''({R})
+ 16\pi G \rho\, ,\\
\label{Tslls26}
0=& \dot H F'({R}) 
 - 3 \left( 4 H^2 \dot H - 4 {\dot H}^2 - 3 H \ddot H 
 - \dddot H \right) F''({R})\nonumber\\
& + 18\left( 4H\dot H + \ddot H\right)^2 F'''({R}) 
+ 4\pi G \left( \rho + p \right) \, .
\end{align}
Now, combining the general equations (\ref{Tslls4}), (\ref{Tslls6}), and 
(\ref{Tslls6B}), 
we obtain
\begin{equation}
\label{Tslls25}
\frac{H}{2\pi} dS = \frac{4\pi}{H^2} \left( \rho + p \right) dt \, .
\end{equation}
Thus, substituting $\rho+p$ from (\ref{Tslls26}) into (\ref{Tslls25})
we acquire the entropy as 
\begin{align}
\label{Tslls27}
S =& - \frac{2\pi}{ G} \int \frac{dt}{H^3}
\Big\{
\dot H F'({R}) + 18\Big( 4H\dot H + \ddot H\Big)^2 F'''({R}) 
\nonumber\\
& - 3 \Big( 4 H^2 \dot H - 4 {\dot H}^2 - 3 H \ddot H - \dddot H \Big) 
F''({R}) \Big\}\, .
\end{align}
Using that ${R}=6 \left( 2 H^2 + \dot H \right)$ 
and $\dot{{R}} = 6\left( 4H \dot H + \ddot H \right)$, the above entropy expression can 
be written as 
\begin{align}
\label{Tslls28}
S =& - \frac{\pi}{ G} \left\{ 
36 \left( \frac{4\dot H}{H^2} \!+ \!\frac{\ddot H}{H^3} \right) F''({R})
 - \left( \frac{1}{H^2} \!-\! \frac{3\dot H}{H^4} \right) \!F'({R}) 
\right.
\nonumber\\
& \left. - \int dt \left[\frac{d}{dt}\left(\frac{3 \dot H}{H^4} \right)
\right] F'({R})\right\} \, .
\end{align}
Note that, as expected, in case of  Einstein gravity $F(R)={R}$, the above 
result reproduces 
the standard result for the entropy, namely
\begin{equation}
\label{Tslls29}
S = \frac{\pi}{ G H^2}=\frac{A}{4G}\, , 
\end{equation}
where the second equality arises using that $r_H=1/H$ and that $A=4\pi r_H^2$.

We focus on the de~Sitter universe, in which $H$ is a constant. In this case 
expression (\ref{Tslls28}) reduces to 
\begin{equation}
\label{Tslls29B}
S = \frac{\pi F'({R})}{ H^2}= \frac{F'({R})A}{4G} \, .
\end{equation}
Interestingly enough, this is the standard result for the entropy in $F(R)$ gravity, 
obtained in \cite{Brevik:2004sd,Cognola:2005de} using the classical 
Euclidean action or the Noether charge method \cite{Iyer:1994ys} in the de~Sitter 
space-time. Hence, recalling that in a de~Sitter universe ${R}=12 H^2$ and that 
$r_H=1/H$, we find that ${R}=12/r_H^2=48\pi/A$. Therefore, if we consider the model where 
$F(R)$ is given by the power of ${R}$, namely
$F(R) \propto {R}^m$, we deduce that (\ref{Tslls29B}) gives
\begin{equation}
\label{Tslls31}
S\propto A^{2-m} \, .
\end{equation}
Thus, making the identification
\begin{equation}
\label{Tslls32}
\delta = 2 - m \, ,
\end{equation}
then the non-extensive Tsallis entropy is reproduced, and we obtain a correspondence 
between $F(R)$ gravity and cosmology from non-extensive thermodynamics. As expected, 
when $F(R)$ gravity becomes general relativity, i.e. for $m=1$, (\ref{Tslls32}) gives 
$\delta=1$ and standard thermodynamics with standard Friedmann equations are reproduced. 
Lastly, we mention here that the above simple correspondence has been shown in the case 
of a de~Sitter universe. For more complicated geometries the correspondence is lost and 
modified cosmology from non-extensive thermodynamics does not have an equivalent $F(R)$ 
gravity description, and hence it corresponds to a novel modification.

\section{Conclusions}
 \label{Conclusions}

In this work we investigated a modified cosmological scenario that arises
from the application of non-extensive thermodynamics with varying exponent in a 
cosmological framework. In particular, it is known that one can apply the 
first law of thermodynamics in the universe apparent horizon and obtain the Friedmann 
equations. Nevertheless, in non-additive systems such as  gravitational ones, the 
usual Bekenstein-Hawking entropy should be replaced by the generalized, non-extensive 
Tsallis entropy. Doing so we obtained modified Friedmann equations, which contain 
new terms quantified by the non-extensive exponent $\delta$. Finally, following quantum 
considerations, we allowed $\delta$ to have a dependence on the scale, with the value 1 
corresponding to standard thermodynamics and standard $\Lambda$CDM cosmology.

Concerning the universe evolution at late times, the new terms that appear due to the 
non-extensive varying exponent constitute an effective dark energy sector. As we showed, 
the universe exhibits the usual thermal history, with the successive sequence of matter 
and dark-energy epochs, and with the transition to acceleration happening around
$z\approx0.6$ in agreement with the observed behavior. Concerning the effective 
dark-energy equation-of-state parameter, we saw that it acquires values close to $-1$ at 
present, while in the past it may lie either in the quintessence or in the phantom 
regime. The interesting feature  is that the above 
behaviors can be obtained even if the explicit cosmological constant is set to zero, 
namely they arise purely from the extra terms in the Friedmann equations.

Confronting the model with SNIa  and $H(z)$ observational data, 
we provided the corresponding contour plots, and we showed that the 
agreement is very good. For the new   
parameters of the scenario at hand we saw that although there is a   tendency for a 
slight  deviation from the $\Lambda$CDM values, $\Lambda$CDM paradigm is included within  
2$\sigma$.

In the early-time universe we saw that the scenario at hand can lead to inflationary 
de~Sitter solutions, which are driven by an effective cosmological constant that includes 
additionally the information of the new terms of non-extensive 
thermodynamics. This feature can provide a
description of both inflation and late-time acceleration with the same parameter choices, 
since the above effective screening gives the necessary enhanced acceleration in 
inflation comparing to dark-energy epoch. This is an advantage comparing to other models 
of the literature, which in general cannot describe inflation and late-time acceleration 
simultaneously since they cannot provide a natural change of the involved parameter 
scales.

Finally, we investigated the correspondence of the scenario at hand with modifications of 
gravity such as the $F(R)$ gravity. As we showed, in the case 
of a de~Sitter universe there is a correspondence between cosmology from non-extensive 
thermodynamics and power-law $F(R)$ gravity, where the non-extensive exponent is related 
to the $F(R)$ exponent. However, for more complicated geometries the correspondence is 
lost and the scenario at hand does not have an equivalent 
$F(R)$ gravity description, forming a novel modification.

In summary, modified cosmology through non-extensive thermodynamics with varying 
exponent is very efficient in describing the universe evolution, from inflation to 
late-time acceleration. It would be interesting to perform further investigations 
on the scenario, such as a joint 
observational analysis using data from Type Ia Supernovae (SNIa), Baryon Acoustic 
Oscillations (BAO), Cosmic Microwave Background (CMB), and Hubble 
parameter observations, or a detailed phase-space analysis in order to extract 
the global features of the 
scenario. These necessary studies lie beyond the scope of this work and are 
left for future works.

\begin{acknowledgments}
The authors would like to thank S. Pan and A. Mukherjee for useful comments.
This article is based upon 
work from CANTATA COST (European Cooperation in Science and Technology) 
action CA15117, 
EU Framework Programme Horizon 2020. 
This work is also supported (in part) by MEXT KAKENHI Grant-in-Aid for 
Scientific Research on Innovative Areas gCosmic Accelerationh No. 15H05890 (S.N.) 
and the JSPS Grant-in-Aid for Scientific Research (C) No. 18K03615 (S.N.), 
and by MINECO (Spain), FIS2016-76363-P, and 
by project 2017 SGR247 (AGAUR, Catalonia) (S.D.O). 
\end{acknowledgments}

\end{document}